\documentclass[aps,pre,twocolumn]{revtex4-2}
\usepackage{amsmath}
\usepackage{amssymb}
\usepackage{graphicx}

\def\t{\vartheta}
\def\e{\varepsilon}
\def\vp{\varphi}
\def\w{\omega}
\def\W{\Omega}
\def\d{\delta}

\begin{document}
\title{Discontinuous transition to synchrony in the Kuramoto-Sakaguchi model with a uniform distribution of frequencies}

\author{A. Pikovsky}
\email{pikovsky@uni-potsdam.de}
 \affiliation{Department of Physics and Astronomy, University of Potsdam, Karl-Liebknecht-Str. 24/25, 14476, Potsdam-Golm, Germany}

\date{\today}

\begin{abstract}
The transition to synchrony in the Kuramoto model of globally coupled phase oscillators with a uniform distribution of natural frequencies is discontinuous. We extend the theory of this transition to the Kuramoto-Sakaguchi model, taking into account a phase shift in coupling. In the thermodynamic limit, we derive dependencies of the order parameters on the coupling strength and the phase shift, and describe two transitions from disorder to partial synchrony and from partial synchrony to complete synchrony. In all cases, the first transition is discontinuous, although for phase shifts close to $\pi/2$, the jump is exponentially small. 
\end{abstract}

\maketitle

\section{Introduction}
Onset of synchronization in a large population of coupled oscillators can be considered as a nonequilibrium disorder-order phase transition~\cite{gupta2018statistical}. As the order parameter, one can use the amplitude of the collective oscillating mode, which vanishes in the disordered state. In a deterministic setup, disorder in the natural frequency of the units plays the role of temperature, and an attractive coupling between oscillators should be strong enough for synchronization to appear. In 1975, Y. Kuramoto suggested his paradigmatic model of globally coupled phase oscillators for this phenomenon, and presented a self-consistent theory of the transition~\cite{Kuramoto-75,Kuramoto-84}. For different extensions of this approach see~\cite{Acebron-etal-05}.

The nature of the transition, i.e., whether it is continuous or discontinuous, heavily depends on the distribution of the natural frequencies of the oscillators. An important special case is the uniform distribution of frequencies. In a seminal paper~\cite{pazo2005thermodynamic}, D. Pazo has shown that the Kuramoto model with a uniform distribution of the frequencies demonstrates, in the thermodynamic limit, a discontinuous transition without hysteresis. Since then, the case of uniform distribution attracted large interest~\cite{GHOSH20133812,yuan2013conformists,Ottino-Strogatz-16,Yagasaki_2025}.

In this paper, we explore synchronization transition for a uniform distribution of natural frequencies in the Kuramoto-Sakaguchi model~\cite{Sakaguchi-Kuramoto-86}, which is an extension of the Kuramoto model for the case of an additional phase shift $\alpha$ in the coupling (see Eq.~\eqref{eq:ks1} below). This phase shift allows for a continuous shift from attractive ($\cos\alpha>0$) to repulsive ($\cos\alpha<0$) coupling, with a conservative coupling ($\alpha=\pm\pi/2$) at the border of attraction and repulsion. One could expect that because coupling is maximally attracting in the Kuramoto case of vanishing $\alpha$, the critical coupling constant grows as one approaches the border of attraction. Indeed, in the case of Cauchy distribution of natural frequencies, one can show analytically~\cite{Ott-Antonsen-08} that the critical coupling diverges $\sim (\cos\alpha)^{-1}$. Interestingly, in the considered below case of a uniform distribution, the threshold of synchronization \textit{decreases} with the parameter of the phase shift $\sim\cos\alpha$. 

The paper is organized as follows. We introduce the model and describe the phenomenology of the synchronization transition in Section~\ref{sec:dt}. In our analysis, we use the method developed by Omelchenko and Wolfrum in Refs.~\cite{omel2012nonuniversal,omel2013bifurcations}; we describe it in Section~\ref{sec:sca}. In Section~\ref{sec:ksc} we present the derivation of the main results. We conclude with a discussion in Section~\ref{sec:con}. 

\section{Discontinuous transition in the Kuramoto-Sakaguchi model}
\label{sec:dt}
The Kuramoto-Sakaguchi model~\cite{Sakaguchi-Kuramoto-86} describes the dynamics of $N$ phase oscillators with a mean-field coupling, and is  formulated as a set of equations 
\begin{equation}
\begin{gathered}
\dot{\vp}_k=\w_k+\frac{\e}{N}\sum_j\sin(\vp_j-\vp_k-\alpha)=\\
\w_k+\e R\sin(\Psi-\vp_k-\alpha)\;.
\end{gathered}
\label{eq:ks1}
\end{equation}
Here $\w_k$ are oscillators' natural frequencies, which are taken from a distribution $g(\w)$.
Collective variables $R$ and $\Psi$ are the amplitude and the argument of the complex mean field
\begin{equation}
Z=R e^{i\Psi}=\frac{1}{N}\sum_j e^{i\vp}\;.
\label{eq:ks2}
\end{equation}
The system has two parameters, the phase shift $\alpha$ and the coupling strength $\e$. The case $\alpha=0$ is the Kuramoto model~\cite{Kuramoto-84}.

Mutual attraction of the phases (which is ensured if $-\pi/2<\alpha<\pi/2$) results in a synchronization transition, which is perfect in the thermodynamic limit $N\to\infty$. In this limit, in the asynchronous state $R=0$, while $R>0$ indicates some synchrony in the population; thus $R$ has been picked by Kuramoto as the order parameter of the nonequilibrium transition synchrony-asynchrony. The nature of the transition heavily depends on the distribution $g(\w)$. For a one-humped distribution with a single smooth maximum, typically a simple second-order continuous transition is observed (see, however, more complex situations for nontrivial distributions in Refs.~\cite{omel2012nonuniversal,omel2013bifurcations}), while for two-humped distributions a first-order transition with a hysteresis is typical.

An interesting intermediate case for the Kuramoto model happens for a uniform distribution of frequencies: here a discontinuous transition to synchrony occurs~\cite{pazo2005thermodynamic}. The goal of this paper is to extend the results of Ref.~\cite{pazo2005thermodynamic} to the Kuramoto-Sakaguchi model. Remarkably, all relevant quantities can be calculated analytically in the thermodynamic limit. Before proceeding with the theory, we describe the phenomenology of the synchronization transition and define its main characteristics.

\begin{figure}
\centering
\includegraphics[width=\columnwidth]{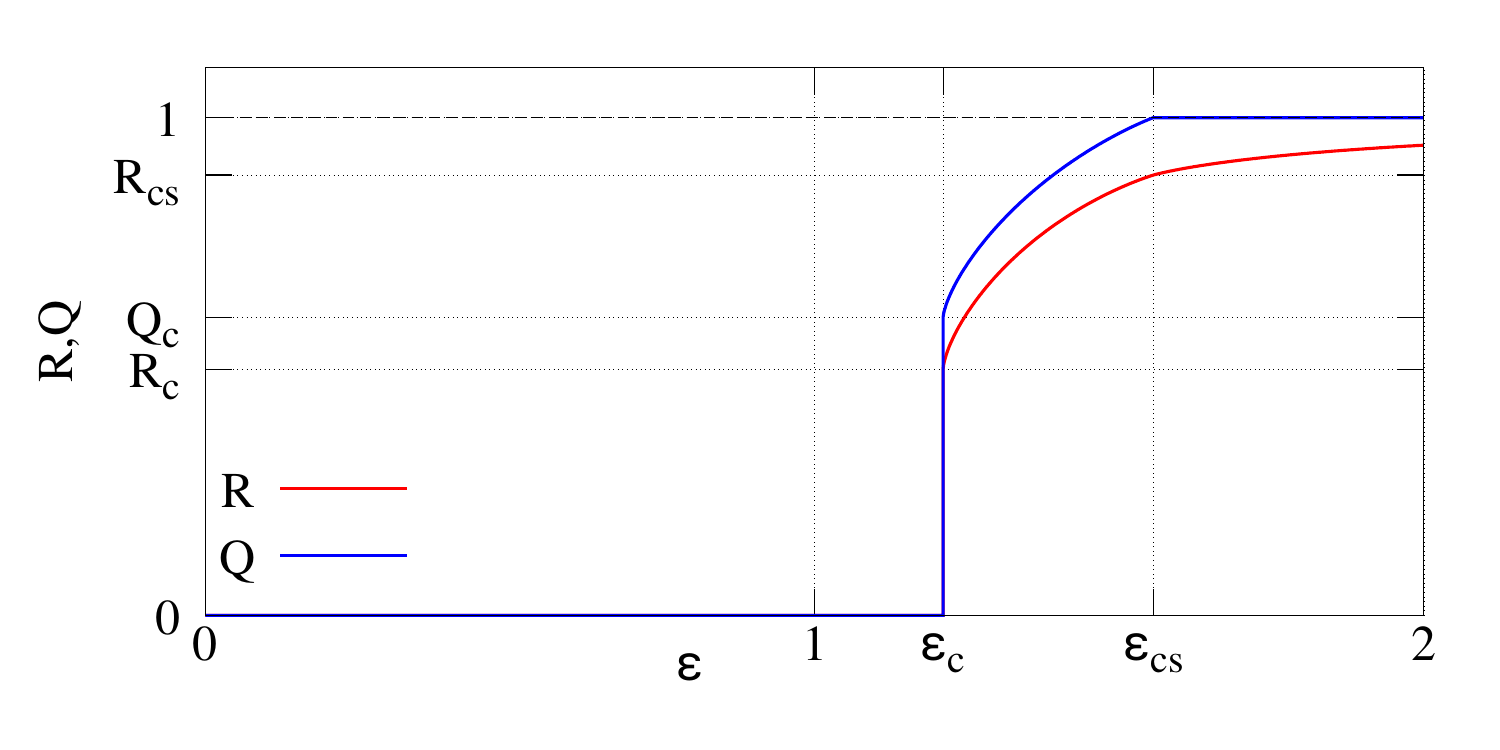}
\caption{Typical dependence of the order parameter $R$ and of the fraction of locked oscillators $Q$ on the coupling strength $\e$ (here drawn for $\alpha=\pi/10$). At $\e_c$, both characteristics jump to values $R_c,Q_c$, demonstrating a discontinuous transition to synchrony. At $\e_{cs}$, the value of $Q$ reaches its maximal value $1$, which marks a transition to complete synchrony.}
\label{fig:rq}
\end{figure}

Typically, one is interested in the dependence of the order parameter $R$ on the coupling strength $\e$ at a fixed phase shift parameter $\alpha$. We show this dependence in Fig.~\ref{fig:rq}. At $\e_c$, there is a discontinuous transition from $0$ to $R_c$; for larger couplings $R$ grows monotonically with $\e$. In this figure, we also show another characteristic of the synchronization transition, the fraction of locked oscillators $Q$. If the mean field amplitude $R$ is non-zero, some oscillators are locked by the mean field (and are also mutually locked), while others have frequencies different from that of the mean field. At the transition point $\e_c$, this fraction also experiences a jump from zero to some value $Q_c$. Further growth of $Q$ leads to another transition at $\e_{cs}$, above which $Q=1$, which means that all units are mutually locked and the dynamics of the ensemble is purely periodic (complete synchrony). Note that here still $R<1$ because the distribution of the phases is not a delta-function.

\begin{figure}
\centering
\includegraphics[width=\columnwidth]{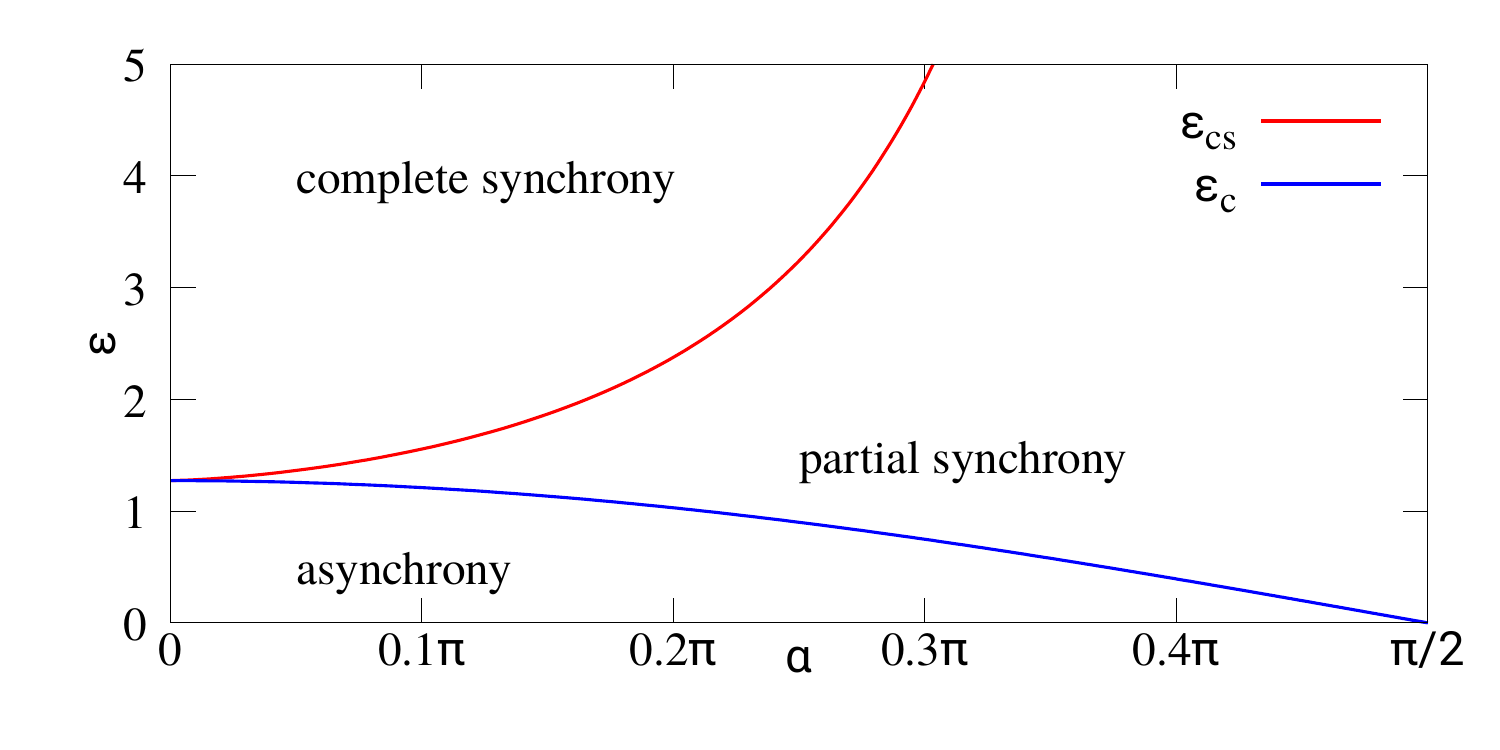}
\caption{Dependence of the two critical values of coupling strength $\e_c$, $\e_{cs}$ on the parameter $\alpha$.}
\label{fig:ec}
\end{figure}

The two critical values of coupling strength, $\e_c$ and $\e_{cs}$, depend on the phase shift parameter $\alpha$ as it is shown in Fig.~\ref{fig:ec}. Note that for $\alpha=0$ $\e_c=\e_{cs}$, in accordance with results of Ref.~\cite{pazo2005thermodynamic}. Figure \ref{fig:ec} can be considered as the full diagram of possible regimes for the model under consideration; the borders are derived analytically below in Section~\ref{sec:ksc}.

\begin{figure}
\centering
\includegraphics[width=\columnwidth]{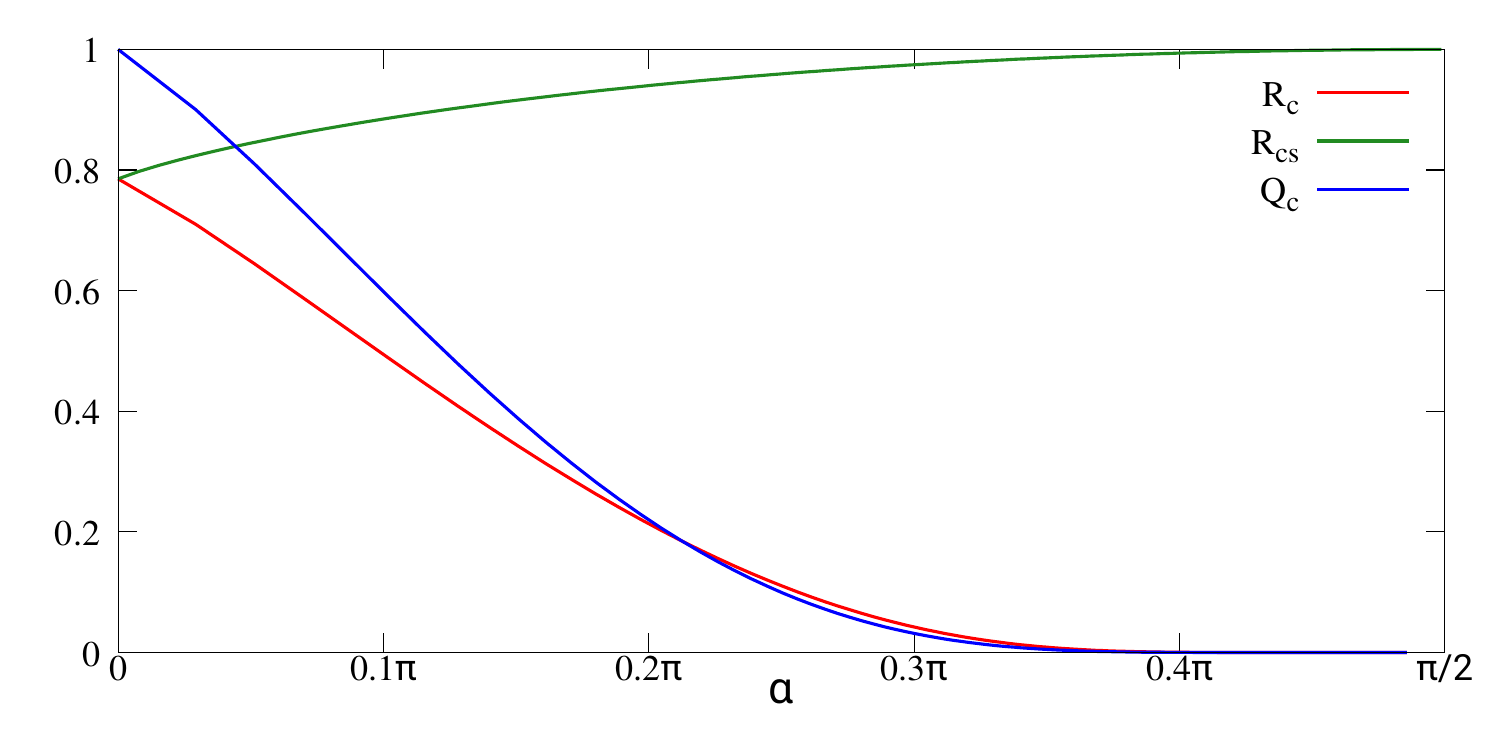}
\caption{Dependencies of the steps $R_c,Q_c$ on the parameter $\alpha$. Additionally, the value $R_{cs}$ (also indicated in Fig.~\ref{fig:rq}) of the order parameter $R$ at the transition to complete synchrony is shown.}
\label{fig:rq2}
\end{figure}

Finally, in Fig.~\ref{fig:rq2}, we show dependencies of $R_c,R_{cs},Q_c$ on $\alpha$. One can see that the jumps $R_c,Q_c$ drastically decrease with $\alpha$, but they are finite in the whole range $0\leq\alpha<\pi/2$, to be discussed in detail below in Section~\ref{sec:smA}.

\section{Self-consistent analysis of the synchronous state} 
\label{sec:sca}

In our analysis, we will follow the approach developed in Refs.~\cite{omel2012nonuniversal,omel2013bifurcations}.
For the sake of consistency, we will succinctly present it here. 

\subsection{Generic case}
In the thermodynamic limit, we denote oscillators with natural frequency $\w$ as $\vp_\w$ and rewrite Eqs.~\eqref{eq:ks1},\eqref{eq:ks2} as
\begin{equation}
\begin{gathered}
\dot{\vp}_\w=\w+\e R\sin(\Psi-\vp_\w-\alpha)\;,\\
R e^{i\Psi}=\langle e^{i\vp}\rangle=\int d\w \, g(\w)\int d\vp\, \rho(\vp|\w) e^{i\vp}\;.
\end{gathered}
\label{eq:ks3}
\end{equation}
Here $\rho(\vp|\w)$ is the density of oscillators with natural frequency $\w$.

One looks for a periodic in time (statistically stationary) solution where $R=const$ and the phase of the mean field rotates with a constant (still unknown) frequency $\W$: $\Psi=\W t+\Psi_0$. It is convenient to make a transformation to the rotating reference frame, where the r.h.s. will be time-independent:
\[
 \vp=\t+\Psi-\alpha=\t+\W t+\Psi_0-\alpha\;.
\]
The physical meaning of the new phase $\t$ is the phase shift with regard to the uniformly rotating phase of the mean field.
Then the equations \eqref{eq:ks3} are rewritten in terms of $\t$ as
\begin{equation}
\begin{gathered}
\dot{\t}_\w=\w-\W-\e R\sin\t_\w\;,\\
R e^{i\alpha}=\langle e^{i\t}\rangle=\int d\w \, g(\w)\int d\t\, \rho(\t|\w) e^{i\t}\;.
\end{gathered}
\label{eq:ks4}
\end{equation}

The next step is to find the stationary distribution $\rho(\t|\w)$ and the integral $\int d\t\, \rho(\t|\w) e^{i\t}$. Let us denote
\begin{equation}
s=\frac{\w-\W}{\e R}
\label{eq:ks5}
\end{equation}
and rewrite \eqref{eq:ks4} as
\begin{equation}
\dot{\t}_\w=\e R(s-\sin\t_\w)\;.
\label{eq:ks6}
\end{equation}
In this equation, the factor $\e R$ defines the time scale and is irrelevant for the stationary invariant distribution of the phase; the latter depends on the parameter $s$ only.

The dynamics of \eqref{eq:ks6} is as follows: (i) for $|s|<1$ there is a stable steady state $\t=\arcsin(s)$; (ii) for $|s|>1$, the phase rotates, and its distribution density is inverse proportional to the velocity of the motion $\sim |s-\sin\t|^{-1}$.
This yields
\begin{equation}
\rho(\t|\w)=\begin{cases}
\delta(\t-\arcsin(s))&\text{for }|s|\leq 1\;,\\ 
\frac{\sqrt{s^2-1}}{2\pi}\frac{1}{|s-\sin\t|} &\text{for } |s|>1\;.
\end{cases}
\label{eq:ks7}
\end{equation}
This allows for calculation of the integral in \eqref{eq:ks4} as follows:
\begin{equation}
\begin{gathered}
\int d\t\, \rho(\t|\w) e^{i\t}=\\
h(s)=\begin{cases}
is(1-\sqrt{1-s^{-2}})&\text{for } |s|>1\;,\\
is+\sqrt{1-s^2} &\text{for } |s|\leq 1\;.
\end{cases}
\end{gathered}
\label{eq:ks8}
\end{equation}
Substitution of this integral in \eqref{eq:ks4} yields the following relation 
\begin{equation}
\begin{gathered}
R e^{i\alpha}=\int d\w \, g(\w)h\left(\frac{\w-\W}{\e R}\right)=\\
\e R \int d s \, g(\W+\e R s)h(s)\;,
\end{gathered}
\label{eq:ks9}
\end{equation}
which has the meaning of a complex self-consistency equation defining the synchronous state.

Let us introduce parameter $A=\e R$ and complex-valued function $f(\W,A)$ as
\begin{equation}
f(\W,A)=\int d s \, g(\W+A s)h(s)\;.
\label{eq:f0}
\end{equation}
With this function, we can find from \eqref{eq:ks9} the parameters $\alpha$ and $\e$ as functions of $\W$ and $A$:
\begin{equation}
R=A|f(\W,A)|,\quad \e=\frac{1}{|f(\W,A)|},\quad \alpha=\text{arg}\,f(\W,A)\;.
\label{eq:ks10}
\end{equation}
This provides a solution to the problem, namely the dependence of the order parameter $R$ on the coupling constant $\e$, in a parametric form.

Additionally, from \eqref{eq:ks8}, one can find which oscillators are locked to the mean field, and which are rotating with respect to the mean field: the locked oscillators have natural frequencies in the interval $\W-A\leq \w\leq \W+A$, the rotating oscillators have natural frequencies outside of this interval.
Thus, the fraction of locked units $Q$ is calculated as 
\begin{equation}
Q=\int_{\W-A}^{\W+A}d\w\;g(\w)\;.
\label{eq:ks11}
\end{equation}
Summarizing, all the quantities of interest are represented in the parametric form via the integrals over the distribution of natural frequencies:
\begin{align}
R&=A|f(\W,A)|,\label{eq:R}\\
\e&=\frac{1}{|f(\W,A)|},\label{eq:e}\\
\alpha&=\text{arg}\,f(\W,A),\label{eq:a}\\
Q&=\int_{\W-A}^{\W+A}d\w\;g(\w), \label{eq:Q}\\
f(\W,A)&=\int d s \, g(\W+A s)h(s).\label{eq:f}
\end{align}

\subsection{Kuramoto model, uniform distribution of frequencies}
We first apply expressions \eqref{eq:R}-\eqref{eq:f} to the Kuramoto model and reproduce results of Ref.~\cite{pazo2005thermodynamic}.
In the Kuramoto model $\alpha=0$. If $g(\w)$ is an even function, then due to symmetry we set $\W=0$ and obtain
\[
f(0,A)=\int_{-1}^{1} g(Ay)\sqrt{1-y^2}\;dy\;.
\]
(here only the even part of $h(s)$ contributes).
For a uniform distribution 
\begin{equation}
g(\w)=\begin{cases} \frac{1}{2}& |\w|<1\;,\\ 0 & |\w|>1\;,\end{cases}
\label{eq:g}
\end{equation}
we obtain
\begin{equation}
f(0,A)=\begin{cases} \frac{\pi}{4} & A\leq 1\;,\\
\frac{1}{2}\left[\frac{1}{A}\sqrt{1-(\frac{1}{A})^2}+\arcsin\frac{1}{A}\right] & A>1\;,\end{cases}
\label{eq:g1}
\end{equation}
and
\begin{equation}
Q=\begin{cases} A & A\leq 1\;,\\
1 & A>1\;.\end{cases}
\label{eq:q1}
\end{equation}
This yields the $R(\e)$ dependence in a parametric form
\begin{equation}
R=Af(0,A),\qquad \e=\frac{1}{f(0,A)}\;.
\label{eq:g2}
\end{equation}
At $\e_c=4/\pi$, the order parameter $R$ takes values $0\leq R\leq \pi/4$, which corresponds to a jump of the order parameter; further at $\e>\e_c$, $R$ grows. At $\e_c$ the value of $Q$ is between $0\leq Q\leq 1$, and for $\e>\e_c$ one has complete synchrony  $Q=1$. The transition to synchrony is discontinuous: at $\e_c$ the values of $R,Q$ jump to $R_c=\pi/4$ and $Q_c=1$, respectively; for $\e>\e_c$ the order parameter $R$ grows and tends to $1$ as $\e\to \infty$.

\section{Kuramoto-Sakaguchi case, uniform distribution of frequencies}
\label{sec:ksc}
Here we consider a general Kuramoto-Sakaguchi case $\alpha\neq 0$ for the uniform distribution \eqref{eq:g}.
\subsection{Analytic representation of the fraction of locked units}
The simplest calculation is that of the fraction of synchronous oscillators $Q$. According to \eqref{eq:Q}, for the uniform distribution of frequencies \eqref{eq:g} we have
\begin{equation}
Q=\frac{1}{2}(\text{min}(\W+A,1)-\text{max}(\W-A,-1))\;.
\end{equation}
This leads to the following representation of the quantity $Q$:
\begin{equation}
Q=\begin{cases}
A& A<1-\W \text{ and } A<1+\W \;,\\
\frac{1+\W+A}{2} & 1+\W<A<1-\W \;,\\
\frac{1+A-\W}{2} & 1-\W<A< 1+\W\;,\\
1& A>1-\W \text{ and } A>1+\W \;.
\end{cases}
\label{eq:Qanal}
\end{equation}

\subsection{Analytic representations of function $f$}
Let us calculate function $f(\W,A)$ (Eq.~\eqref{eq:f}) in the case of a uniform distribution \eqref{eq:g}:
\begin{equation}
f(\W,A)=\int d s \, g(\W+A s)h(s)=\frac{1}{2}\int_{-(1+\W)/A}^{(1-\W)/A} h(s)\,ds\;.
\label{eq:f1}
\end{equation}
It is convenient to separate real and imaginary parts of $h(s)$ (Eq.~\eqref{eq:ks8})
\begin{gather}
h(s)=h_r(s)+i h_i(s), \label{eq:hall}\\
h_r(s)=\begin{cases}
0& |s|>1\;,\\
\sqrt{1-s^2} & |s|\leq 1\;,
\end{cases} \label{eq:hr}\\
h_i(s)=\begin{cases}
s(1-\sqrt{1-s^{-2}})& |s|>1 \;,\label{eq:hi}\\
s& |s|\leq 1\;.
\end{cases}
\end{gather}
Taking indefinite integrals of these expressions, we obtain the following functions $H_r,H_i$:
\begin{equation}
\begin{gathered}
H_r(x)=\int^x h_r(s)\,ds=\\
\begin{cases} -\frac{\pi}{4} & x<-1\;,\\
\frac{1}{2}(\arcsin x+x\sqrt{1-x^2})& -1\leq x\leq 1\;,\\
\frac{\pi}{4}& x>1\;.
\end{cases}\\
H_i(x)=\int^x h_r(s)\,ds=\\
\begin{cases} \frac{x^2}{2} & |x|\leq 1\;,\\
\frac{1}{4}\left[2 x^2(1-\sqrt{1-x^{-2}})+\ln\frac{1+\sqrt{1-x^{-2}}}{1-\sqrt{1-x^{-2}}}\right] & |x|>1\;.
\end{cases}
\end{gathered}
\label{eq:inth}
\end{equation}
Substituting these functions in \eqref{eq:f1}, we obtain an analytic expression of funcion $f$:
\begin{equation}
\begin{gathered}
f(\W,A)=\frac{1}{2}\left[H_r\left(\frac{1-\W}{A}\right)-H_r\left(-\frac{1+\W}{A}\right)\right]+\\
i\frac{1}{2}\left[H_i\left(\frac{1-\W}{A}\right)-H_i\left(-\frac{1+\W}{A}\right)\right]\;.
\end{gathered}
\label{eq:f2}
\end{equation}

Below, we need to consider the limit of a small $A$. For this purpose, it is convenient
to represent the functions $H_r,H_i$ from \eqref{eq:inth} as functions of $y=1/x$.

For the real part, we have
\begin{equation}
\begin{gathered}
G_r(y)=H_r\left(\frac{1}{y}\right)=\\
\begin{cases} -\frac{\pi}{4} & -1<y<0\;,\\
\frac{1}{2}(\arcsin (y^{-1})+y^{-1}\sqrt{1-y^{-2}})& |y|\geq 1\;,\\
\frac{\pi}{4}& 0<y<1\;.
\end{cases}
\end{gathered}
\label{eq:rp}
\end{equation}
For the imaginary part, we have 
\begin{equation}
\begin{gathered}
G_i(y)=H_i\left(\frac{1}{y}\right)=\\
\begin{cases}
\frac{1}{2y^2}& |y|\geq 1\;,\\
\frac{1}{4}\left[2 y^{-2}(1-\sqrt{1-y^{2}})+\ln\frac{1+\sqrt{1-y^{2}}}{1-\sqrt{1-y^{2}}}\right]& |y|\leq 1\;.
\end{cases}
\end{gathered}
\label{eq:ip0}
\end{equation}
A slightly different equivalent form of $G_i$ is obtained by multiplying and dividing by $1+\sqrt{1-y^{2}}$:
\begin{equation}
\begin{gathered}
G_i(y)=\\
\begin{cases}
\frac{1}{2y^2}& |y|>1\;,\\
\frac{1}{2}\frac{1}{1+\sqrt{1-y^2}}+\frac{1}{4}\ln(2+2\sqrt{1-y^2}-y^2)-\frac{1}{2}\ln|y| & |y|\leq 1\;.
\end{cases}
\end{gathered}
\label{eq:ip}
\end{equation}
In terms of these functions, the function $f$ is represented as
\begin{equation}
\begin{gathered}
f(\W,A)=\frac{1}{2}\left[G_r\left(\frac{A}{1-\W}\right)-G_r\left(-\frac{A}{1+\W}\right)\right]+\\
i\frac{1}{2}\left[G_i\left(\frac{A}{1-\W}\right)-G_i\left(-\frac{A}{1+\W}\right)\right]\;.
\end{gathered}
\label{eq:f22}
\end{equation}

The obtained expressions for the function $f$ \eqref{eq:f2} and \eqref{eq:f22} can be substituted in Eqs.~\eqref{eq:R}-\eqref{eq:a} to obtain an analytic dependence of the order parameter $R$ on coupling strength $\e$ and phase shift $\alpha$. The disadvantage of this representation is that by choosing values of $A$ and $\W$, we cannot predict to which phase shift $\alpha$ these values correspond. To have a dependence $R(\e)$ at a prescribed value of the phase shift $\alpha$ like one depicted in Fig.~\ref{fig:rq}, we for each value of $A$ considered many different values of $\W$, calculated corresponding phase shifts, and determined the proper $\W$ via a one-dimensional root search algorithm. In this way, Figure~\ref{fig:rq} was created.

\subsection{Critical coupling: Limit of vanishing amplitude $A$}
Let us look at the limit $A\to 0$, this, according to \eqref{eq:R}, corresponds to $R=0$, thus to the critical coupling $\e_c$. In this limit, it is convenient to consider representations \eqref{eq:rp},\eqref{eq:ip}.

For small $A\to 0$ we obtain from \eqref{eq:rp}
\begin{equation}
\begin{gathered}
f_r(\W,0)=\lim_{A\to 0}\frac{1}{2}\left[G_r\left(\frac{A}{1-\W}\right)- G_r\left(\frac{A}{-1-\W}\right)\right]=\\
\begin{cases} \frac{\pi}{4} & |\W|<1\;,\\ 0 & |\W|>1\;.
\end{cases}
\end{gathered}
\label{eq:fre}
\end{equation}
For the imaginary part we obtain from \eqref{eq:ip} for  $A\to 0$
\begin{equation}
\begin{gathered}
f_i(\W,0)=\lim_{A\to 0}\frac{1}{2}\left[G_i\left(\frac{A}{1-\W}\right)- G_i\left(\frac{A}{-1-\W}\right)\right]=\\
\frac{1}{4}\ln\left|\frac{1-\W}{1+\W}\right|\;.
\end{gathered}
\label{eq:fim}
\end{equation}
Functions $f_i(\W,0),f_r(\W,0)$ yield the critical line on the $(\e,\alpha)$ plane. Because
according to \eqref{eq:e},\eqref{eq:a}
\[
\e_c^2=\frac{1}{f_i^2+f_r^2},\qquad \alpha=\text{arg}(f_r+if_i),
\]
we obtain an explicit expression for the critical coupling
\begin{equation}
\e_c^2=\frac{1}{f_r^2}\frac{1}{1+\tan^2\alpha}=\frac{16\cos^2\alpha}{\pi^2}\quad \Rightarrow\quad \e_c=\frac{4}{\pi}\cos\alpha\;.
\label{eq:ec}
\end{equation}
We stress that because $f_r\geq 0$, the 
possible range of values of parameter $\alpha$ is $-\pi/2< \alpha <\pi/2$. In this range, the coupling is attractive, while for $\alpha>\pi/2$ and $\alpha<-\pi/2$ the coupling is repulsive and there is no synchronization transition.

A remarkable property of expression \eqref{eq:ec} is that the synchronization threshold is maximal for maximal attraction $\alpha=0$ and vanishes for neutral (conservative) coupling $\alpha=\pm\pi/2$. This is in contrast to the critical coupling behavior for other typical distributions of the frequencies $g(\w)$ considered in the literature. For example, for the Cauchy distribution of natural frequencies, the critical coupling strength is \textit{inverse} proportional to $\cos\alpha$ and tends to infinity in the conservative limit.

\subsection{Small amplitudes $A$}
\label{sec:smA}

\begin{figure}[!htb]
\centering
\includegraphics[width=0.7\columnwidth]{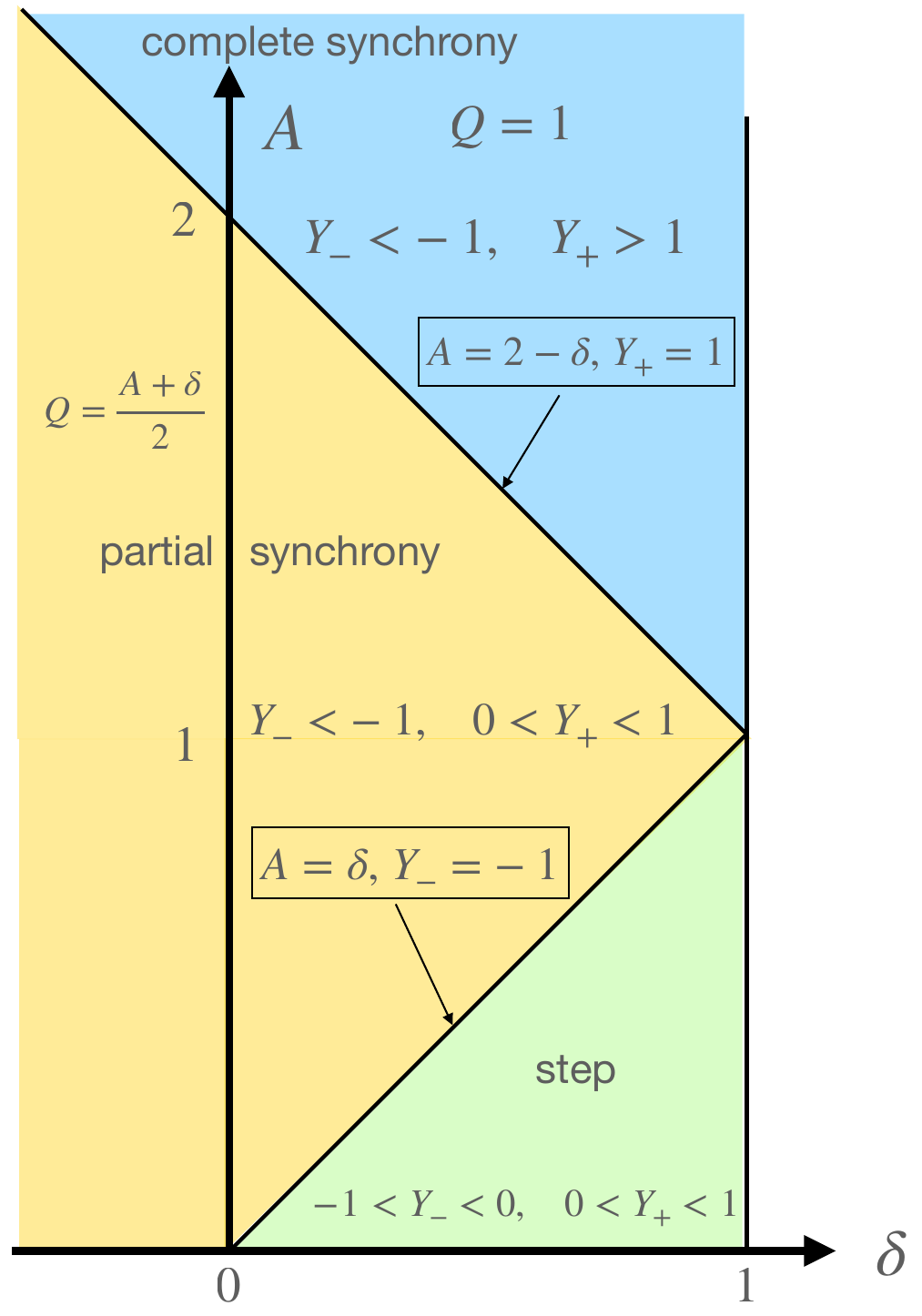}
\caption{Regions of different regimes on the ($A,\d$) plane, and the corresponding values of $Y_{\pm},Q$. }
\label{fig:sk}
\end{figure}

As it follows from \eqref{eq:fim}, the case $\pi/2>\alpha >0$ corresponds to $\infty>f_i>0$ and thus to $-1<\W<0$ at $A=0$.
Without loss of generality, we consider negative values of parameter $\W$ also for $A>0$ (the case of negative $\alpha$ corresponds to the symmetric case of positive frequencies $\W$). To simplify the analysis below, we define $\d=1+\W$, $\d\leq 1$. In Eqs. \eqref{eq:fim},\eqref{eq:fre} enter values
\[
\frac{A}{-1-\W}=-\frac{A}{\d}=Y_-\;,\qquad \frac{A}{1-\W}=\frac{A}{2-\d}=Y_+\;.
\]
The regions of different values of $Y_-,Y_+$ in dependence on $A,\d$ are depicted in Fig.~\ref{fig:sk}.
In these notations, quantity $Q$ (from \eqref{eq:Qanal}) is represented as:
\begin{equation}
Q=\begin{cases} 1 & A>2-\delta\;,
\\ \frac{A+\delta}{2}& \delta\leq A\leq 2-\delta\;,\\
A &A<\delta\;.
\end{cases}
\label{eq:Qdanal}
\end{equation}
These values are also depicted in Fig.~\ref{fig:sk}.

Here we consider small values of $A$, so that $0\leq A\lesssim 1,\;\d\leq 1$. 
In terms of the variables $Y_-,Y_+$ we rewrite Eq.~\eqref{eq:f22} as
\begin{equation}
f(\W,A)=\frac{1}{2}\left[G_r(Y_+)-G_r(Y_-)]+i\frac{1}{2}[G_i(Y_+)-G_i(Y_-)\right]\;.
\label{eq:f23}
\end{equation}
For small $A$ we have  $0\leq Y_+\leq 1$, while with regard to $Y_-$ we have two domains:
\[
\begin{cases} -1<Y_-<0 & \text{ for } A<\d\;,\\
Y_-<-1 & \text{ for } A>\d\;.
\end{cases}
\]
Thus the function $f=f_r+if_i$ has two representations in these domains:\\
For $A<\d$:
\begin{equation}
\begin{gathered}
f_r(\W,A)=F_r(Y_-,Y_+)=\frac{\pi}{4}\\
f_i(\W,A)=F_i(Y_-,Y_+)=\\
\frac{1}{2}\left[\frac{1}{2}\frac{1}{1+\sqrt{1-Y_+^2}}+\frac{1}{4}\ln(2+2\sqrt{1-Y_+^2}-Y_+^2)-\frac{1}{2}\ln|Y_+|\right]-\\
\frac{1}{2}\left[\frac{1}{2}\frac{1}{1+\sqrt{1-Y_-^2}}+\frac{1}{4}\ln(2+2\sqrt{1-Y_-^2}-Y_-^2)-\frac{1}{2}\ln|Y_-|\right]\;.
\end{gathered}
\label{eq:f31}
\end{equation}
For $A>\d$:
\begin{equation}
\begin{gathered}
f_r(\W,A)=F_r(Y_-,Y_+)=\\
\frac{\pi}{8}-\frac{1}{4}(\arcsin (Y_-^{-1})+Y_-^{-1}\sqrt{1-Y_-^{-2}})\\
f_i(\W,A)=F_i(Y_-,Y_+)=\\
\frac{1}{2}\Big[\frac{1}{2}\frac{1}{1+\sqrt{1-Y_+^2}}+\frac{1}{4}\ln(2+2\sqrt{1-Y_+^2}-Y_+^2)-\\
\frac{1}{2}\ln|Y_+|\Big]-\frac{1}{4}\frac{1}{Y_-^2}\;.
\end{gathered}
\label{eq:f32}
\end{equation}

Expressions \eqref{eq:f31},\eqref{eq:f32} allow for the analysis of the transition to synchrony at $\e_c$.

In the domain $A<\d$, because $f_r=const=\pi/4$, the value of $f_i=f_r\tan\alpha$ is also constant for fixed $\alpha$. Thus, in this domain $\e=\text{const}$ while $A$ varies from $0$ at $Y_-=0$ to $\d$ at $Y_-=-1$. This domain corresponds to a step in the $R$ vs $\e$ dependence and is correspondingly marked in Fig \ref{fig:sk}.

If $Y_-=-1$ then $Y_+=\frac{\d}{2-\d}$. Thus, the step amplitude $R$ is determined in parametric form ($0<\d<1$ is the parameter) from the equations
\begin{equation}
\begin{gathered}
\frac{\pi}{4}\tan\alpha=F_i\left(-1,\frac{\d}{2-\d}\right)=\\
=\frac{1}{4}
\left(\frac{1}{1+\sqrt{1-\d^2/(2-\d)^2}}-1\right)+\\
\frac{1}{8}\ln\left(2+2\sqrt{1-\d^2/(2-\d)^2}-\d^2/(2-\d)^2\right)-\\
\frac{1}{4}\ln\frac{\d}{2-\d}\;,\\
R_c=\d\sqrt{(\pi/4)^2+F_i^2\left(-1,\frac{\d}{2-\d}\right)}\;,\\
Q_c=\d\;.
\end{gathered}
\label{eq:step}
\end{equation}
Here we added the value of $Q$ that follows from \eqref{eq:Qdanal}.
Expression \eqref{eq:step} gives the value of the steps $R_c,Q_c$ as functions of phase shift $\alpha$, in a parametric form. Figure \ref{fig:rq2} is drawn using this formula.

Let us look at the limit $\alpha\approx \pi/2$. It corresponds to $0<\d\ll 1$. In this limit the main contribution to $F_i$ is from the last logarithm term: $F_i\approx -\frac{1}{4}\ln(\d/2) \gg 1$. This allows for representing $\d\approx 2\exp(-\pi\tan\alpha)$ so that
\begin{equation}
R_c\approx \frac{\pi}{2}\exp(-\pi\tan\alpha)\tan\alpha,\quad Q_c\approx 2\exp(-\pi\tan\alpha)\;.
\label{eq:step1}
\end{equation}

\begin{figure}[!htb]
\includegraphics[width=\columnwidth]{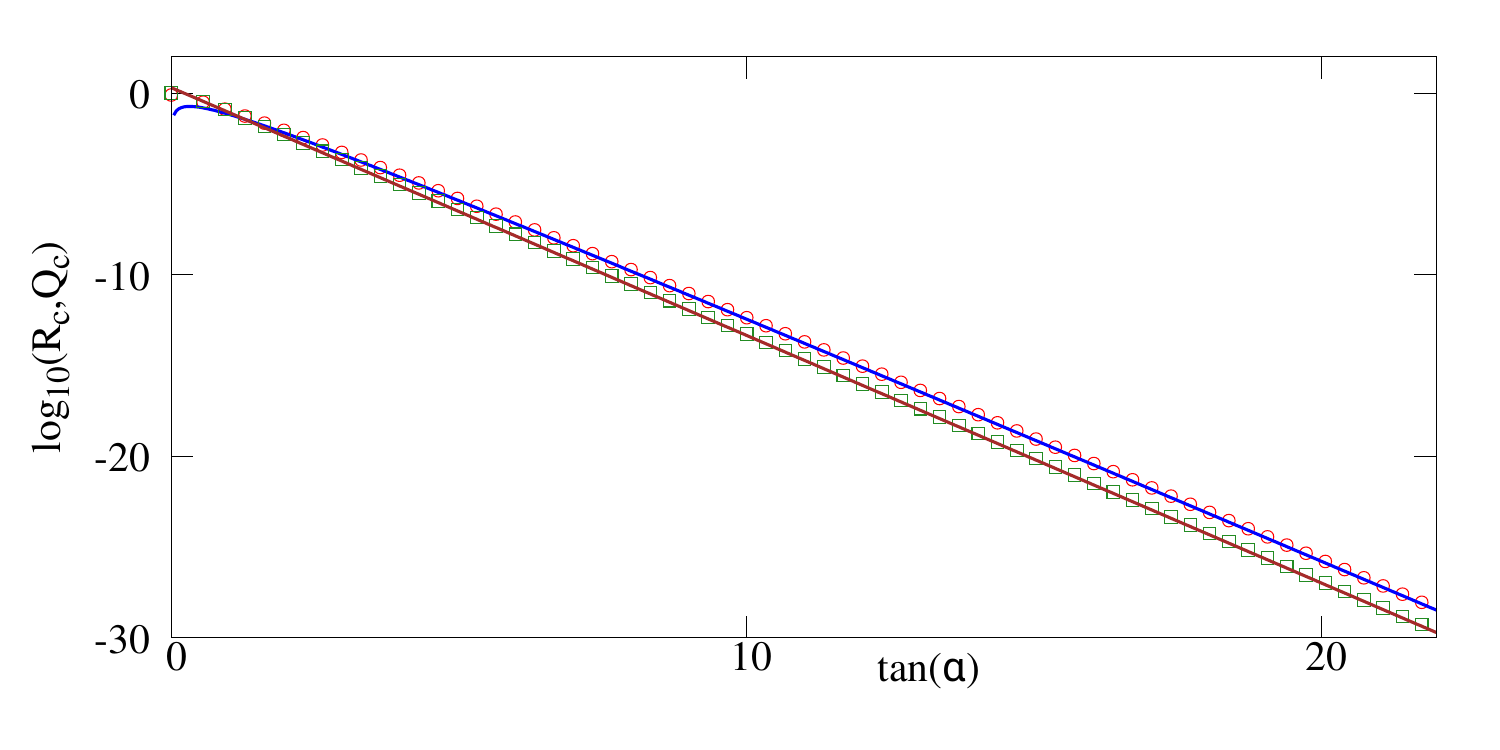}
\caption{Parameters of the discontinuous transition $R_c,Q_c$ as functions of $\alpha$. The values of $R_c$ and $Q_c$ according to exact expressions~\eqref{eq:step} are shown with red circles and green squares, respectively.  The blue and the brown lines show approximate expressions \eqref{eq:step1}. }
\label{fig:step}
\end{figure}

Figure \ref{fig:step} shows the steps in $R,Q$ as functions of $\alpha$ and the comparison to the approximate formula \eqref{eq:step1}. Remarkably, the steps remain finite in the whole range of parameter $\alpha$, albeit for $\alpha$ close to $\pi/2$ they are exponentially small. 

\subsection{Transition to complete synchrony}
\begin{figure}[!htb]
\includegraphics[width=\columnwidth]{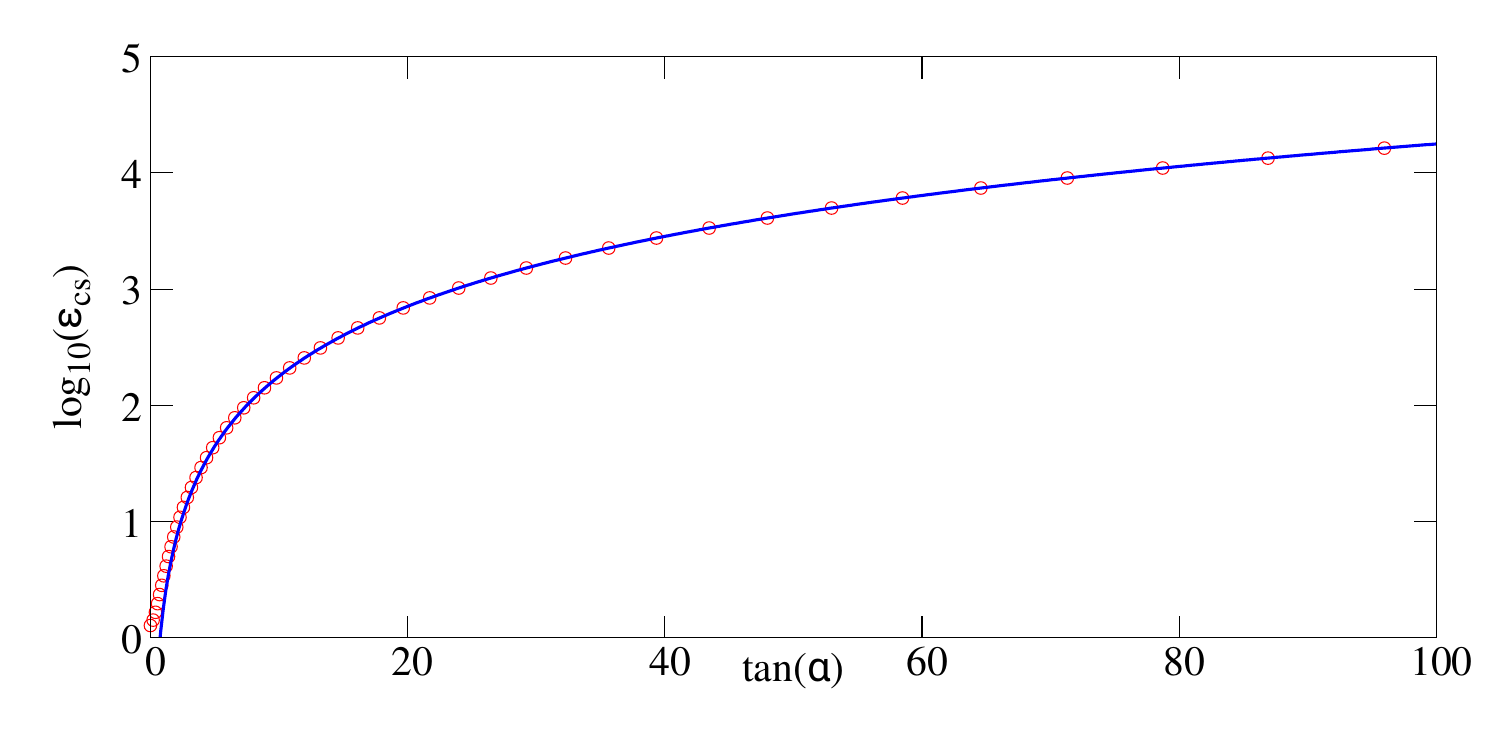}\hfill
\caption{Comparison of the exact values of $\e_{cs}$ calculated according to \eqref{eq:cs} (circles) with the asymptotic at $\alpha\to\pi/2$ expression \eqref{eq:cs2} (blue line).}
\label{fig:cs}
\end{figure}

At complete synchrony, all oscillators are locked and $Q=1$. This happens, according to expression \eqref{eq:Qdanal}, on the line $A=2-\d$, see Fig.~\ref{fig:sk}. At these values of $A,\d$ we have
$Y_+=1$ and $Y_-=\frac{\d-2}{\d}$.
Thus, according to general expressions \eqref{eq:rp},\eqref{eq:ip},\eqref{eq:f22} we have
\begin{gather*}
G_r(Y_+)=\frac{\pi}{4},\qquad G_i(Y_+)=\frac{1}{2},\\
G_r(Y_-)=\frac{1}{2}\arcsin(1/Y_-)+\frac{1}{2}\frac{\sqrt{1-(1/Y_-)^2}}{Y_-}=\\
\frac{1}{2}\arcsin\frac{\d}{\d-2}-\sqrt{1-\d}\frac{\d}{(\d-2)^2}\;,\\
G_i(Y_-)=\frac{\d^2}{2(\d-2)^2}\;,\\
f_r^{cs}=\frac{1}{2}\left(\pi/4-\frac{1}{2}\arcsin\frac{\d}{\d-2}+\sqrt{1-\d}\frac{\d}{(\d-2)^2}\right)\;,\\
f_i^{cs}=\frac{1}{2}\left(\frac{1}{2}-\frac{\d^2}{2(\d-2)^2}\right)=\frac{1-\d}{(\d-2)^2}\;.
\end{gather*}
According to these expressions, the border of complete synchrony $\e_{cs}$ as a function of $\alpha$ is given parametrically
\begin{equation}
\begin{gathered}
\alpha=\text{arg}(f_r^{cs}(\d)+if_i^{cs}(\d)),\quad \e_{cs}=\frac{1}{\sqrt{(f_r^{cs})^2(\d)+(f_i^{cs})^2(\d)}}\;,\\
R_{cs}=A/\e=(2-\d)\sqrt{(f_r^{cs})^2(\d)+(f_i^{cs})^2(\d)}\;.
\end{gathered}
\label{eq:cs}
\end{equation}
These expressions are used to draw the curve $\e_{cs}(\alpha)$ in Fig.~\ref{fig:cs}.

Next, we explore the rate of divergence of the value of $\e_{cs}$ at $\alpha\to\pi/2$.
To take limit of large negative $\d$ where $\alpha\to\pi/2$, it is more convenient to represent $f_r$ via expressions \eqref{eq:f22} (using $X_+=1/Y_+=1,\;X_-=1/Y_-=\d/(\d-2)$):
\begin{gather*}
f_r^{cs}=\frac{1}{2}[H_r(X_+)-H_r(X_-)]=\\
\frac{1}{2}\int_{\d/(2-\d)}^1\sqrt{1-x^2}\;dx\approx \frac{4}{3}(2-\d)^{-3/2}
\end{gather*}
For $\d\to\infty$ we have $f_r^{cs}\approx \frac{4}{3}|\d|^{-3/2}$ and $f_i^{cs}\approx \d^{-1}$. Thus $\tan\alpha=f_i^{cs}/f_r^{cs}\approx \frac{3}{4}|\d|^{1/2}$ and 
\begin{equation}
\e_{cs}\approx |\d|\approx \frac{16}{9}\tan^2\alpha.
\label{eq:cs2}
\end{equation}
In Fig.~\ref{fig:cs} we compare the exact value of threshold of the complete synchrony $\e_{cs}$ vs $\alpha$ according to \eqref{eq:cs} with the approximation \eqref{eq:cs2}.

\section{Conclusions}
\label{sec:con}

In this paper, we presented an analytic theory of the synchronization transition in the Kuramoto-Sakaguchi model of coupled phase oscillators, for a uniform distribution of natural frequencies. Because the distribution has a finite support, two essential transitions can be identified: (i) First, at $\e_c$, a discontinuous transition from disorder to partial synchrony happens, at which a fraction of oscillators becomes synchronized (mutually and with the mean field), while other units are not locked. (ii) The second transition at $\e_{cs}\geq \e_c$ is from the partially synchronous state to complete synchrony, at which all oscillators are locked to the mean field. This second transition does not occur if the distribution of frequencies has infinite support (like a Gaussian or Cauchy distribution). 

The two critical coupling strengths depend on the phase shift in the coupling. They coincide at the vanishing phase shift (Kuramoto model), so there is just one transition from disorder to complete synchrony. Remarkably, the first critical coupling $\e_c$ \textit{decreases} with the phase shift and tends to zero as the coupling becomes conservative at $\alpha=\pm\pi/2$. This property is in contradistinction to a popular case of a Cauchy distribution of frequencies, where the critical coupling \textit{increases} when the coupling becomes less attractive. Notably, although the critical coupling for $\alpha$ close to $\pm\pi/2$ is small, the appearing synchrony is very weak because the discontinuous step of the order parameter is exponentially small in this limit.  Moreover, the range of partial synchrony also diverges close to the neutral coupling, and the second critical value $\e_{cs}$ diverges $\sim\tan^2\alpha$ in this limit. The question of whether these features can be found in other distributions with finite support, such as a beta distribution, is an interesting problem for future study.

\acknowledgements
The author thanks F. Bagnoli and S. Iubini for fruitful discussions.

%

\end{document}